\documentclass[12pt]{article}
\begin{document}
\newcommand{\be}{\begin{equation}}
\newcommand{\ee}{\end{equation}}
\newcommand{\ba}{\begin{array}}
\newcommand{\ea}{\end{array}}
\newcommand{\bc}{\begin{center}}
\newcommand{\ec}{\end{center}}
\newcommand{\beq}{\begin{eqnarray}}
\newcommand{\eq}{\end{eqnarray}}
\def \l {\label}
\bc
{\large{\bf{Baxters's $Q$-operators for the simplest $q$-deformed
model.}}}

\vspace*{10mm}

{\bf A.E. Kovalsky}\\
{\it MIPT, Dolgoprudny, Moscow reg., \\
 IHEP, Protvino, Moscow reg., Russia,} \\

{\bf G.P. Pronko}\\
{\it IHEP, Protvino, Moscow reg., Russia,\\
International Solvay Institute, Brussels, Belgium}

\ec

\begin{abstract}
In the present paper we describe the procedure of the $Q$ - operators
construction for the $q$ - deformed model, described by the Lax operator,
which is important to  formulate the Bethe ansatz for the Sin - Gordon
model. This Lax operator can also be considered as some massless limit of
the Lax operator of SG model. We constructed two $R$ operators which are
the universal intertwiners for the Lax operators. The traces of its
monodromies over the auxiliary space are Baxter operators i.e. the
operator solutions of $T-Q$ equation. We also found the intertwining
relations which imply the mutual commutativity of the corresponding $Q$ -
operators. 
\end{abstract}
 
\section{Introduction}
This article continues the series of articles devoted to the investigation
of Baxter $Q(x)$ operators for various quantum integrable systems. Long
ago, in his famous papers \cite{Baxter} Baxter has introduced this object
for the solution of the eigenvalue problem of $XYZ$ spin chain. In the
frameworks of the Quantum Inverse Scattering Method or Algebraic Bethe
Ansatz (see, for example, \cite{Faddeev}) $Q$ operator could be defined as
follows. Let $T(x)=L_N(x) \dots L_1(x)$ be the momodromy matrix of the Lax
operator $L_k(x)$. The trace of the monodromy is the transfer matrix
$t(x)$. In the cases, we are going to consider the Lax operators, as well,
as its monodromy $T(x)$ are intertwined by  rational or trigonometric $R$
- matrices. Then $Q(x)$ operator is defined by the following equations:
\be
t(x)Q(x)=a(x)Q(x+i)+b(x)Q(x-i)
\ee
in the rational case and
\be
t(x)Q(x)=a(x)Q(qx)+b(x)Q(q^{-1}x)
\l{b1}
\ee
in the trigonometric case. As it follows from the general approach
\cite{Faddeev}, the roots of the eigenvalues of $Q$ are the solutions of
Bethe
equation and consequently it defines the eigenvectors and
eigenvalues of $t(x)$. Functions $a(x)$ and $b(x)$ are in our case
polynomials (usual or trigonometric) and determined by the factorization
of the quantum determinant of the Lax operator. These equations are
discrete analogues of the differential equations of the second order (see
for example \cite{Gaudin}, where was shown how Baxter equation for
inhomogeneous XXX spin chain turns to the differential equation of the
second order, which allows to find spectrum of the Gaudin model). So we
expect two independent solution $Q$ of Baxter equation with the same $t$.
Indeed, in the paper \cite{PrStr}, where authors considered the case of
XXX and XXZ spin chains the second polynomial solution was found (this
fact was first pointed out in \cite{BLZ}). In the present paper we
consider the problem of construction of operator $Q$ i.e. the  operator
solution of (\ref{b1}) with given transfer matrix $t(x)=tr~T(x)$. Actually
the method presented in our paper is similar to the method which Baxter
exploited for quantization of XXZ and XYZ spin chains \cite{Baxter}. In
the previous papers \cite{TodaPronko,DST,q-Toda} (see also
\cite{DSTSklyanin} where was noticed the relation of $Q$ - operators with
quantum Backlund transformation) the expressions for two basic operators
$M$ acting in quantum spaces and auxiliary infinite-dimensional space were
derived. In the paper \cite{TodaPronko} also was discovered the
relationship between $Q$ - operator and the Bloch solutions of quantum
linear problem. The traces of the monodromies of $M$ - operators are the
$Q$ - operators. In the paper \cite{XXXPronko} the one-parametric family
of $Q$ - operators for the XXX spin chain was constructed, however there
is the problem to construct two independent solution in this case. In the
present paper we have constructed the basic $M$ - operators for the
simplest Lax operator connected with the trigonometric $R$  - matrix. 

\section{The intertwining relations with the Lax operator}
\subsection{Universal $R$ - matrix as $Q$ - operator}
Let us consider the problem of construction of Baxter operators for the
integrable model connected with the Lax operator ($x$ is the spectral
parameter):
\be
L(x)=\left(\ba{cc}
u & x v^{-1} \\ -x v & u^{-1} \ea \right) 
\l{Lax}
\ee
Operators $u$ and $v$ form the Weyl pair:
\be
uv=qvu.
\ee
Here we consider the case of $|q|<1$. We will use not only operators
generated by the integer powers of $u$ and $v$ but also the noninteger
powers of $u$ and $v$. So the operators $u$ and $v$ have to be considered
as the exponents of Heisenberg operators $p$ and $q$ with the commutation
relation $[p,q]=-i$.
It is well known that the operators $L$ are intertwined by the
trigonometric $R$ matrix:
\be
R(z)=\left( \ba{cccc} qz-q^{-1}z^{-1} & & & \\ & z-z^{-1} & q-q^{-1} & \\
& q-q^{-1} & z-z^{-1} \\
& & & qz-q^{-1}z^{-1} \ea \right)
\l{trig}
\ee
$R$ - matrix acts in the tensor product of the two dimensional auxiliary
spaces 1 and 2, $L_1$ and $L_2$ act in local quantum space and in the
spaces 1 and 2 respectively. The intertwining equations are
\be
R_{12}(x/y) L_1(x) L_2(y)=L_2(y) L_1(x) R_{12}(x/y)
\l{two}
\ee
Since the operators $u$ and $v$ are invertible there is no usual Bethe
anzats for the monodromy matrix of Lax operators (\ref{Lax}). However it
is possible to construct Bethe ansatz for the product of two Lax operators
and to arrive to the formulation of the Sin-Gordon model within the QISM
approach (see, for example \cite{Faddeev}). 

First we construct the universal $R$ - matrix.  The defining equations
are:
\be
R_{12} L_1(x)L_2(y)=L_2(y)L_1(x)R_{12}.
\l{univ}
\ee
In contrast to (\ref{two}), $R_{12}$ intertwines $L_{1,2}$ in ''quantum''
space.
Here indices 1 and 2 denote different Weyl spaces (with the same $q$) and
Lax operators act in the common auxiliary two-dimensional space. The
corresponding four equations are:
\beq
&&R\cdot(u_1u_2-xy \cdot v_2v_1^{-1})=(u_1u_2-xy \cdot v_1v_2^{-1}) \cdot
R \nonumber\\
&&R \cdot (u_1^{-1}u_2^{-1}-xy \cdot v_1v_2^{-1})=
(u_1^{-1}u_2^{-1}-xy \cdot v_2v_1^{-1}) \cdot R \nonumber\\
&&R \cdot (y u_1v_2^{-1}+xu_2v_1)=(x u_1v_2^{-1}+yu_2v_1) \cdot R
\nonumber\\
&&R\cdot (yu_1^{-1}v_2+xu_2v_1)=(yu_1 v_2+xu_2^{-1}v_1) \cdot R
\eq
It is clear that $R=R(x/y)$. Now let us use the following ansatz:
\be
R_{12}(z)=P_{12}r_{12}
\ee
where $P_{12}$ is the permutation operator i.e.:
\be
u(v)_{1,2}\cdot P_{12}=P_{12} \cdot u(v)_{2,1}
\ee
This ansatz leads to the following single equation (along with $[r,u_1
u_2]=0$):
\be
r(z)\left\{qw_{12}+z\right\}u_2^{-1}v_1^{-1}=u_2^{-1}v_1^{-1}\left\{1+q^{-
1}w_{12}\right\}r(z)
\ee
We choose the following solution of this equation:
\be
r_{12}(z)=f(u_1 u_2,z)g(w_{12},z)
\ee
where $f(z)=(u_1u_2)^{lnz/lnq}$, $w_{12}=u_1v_1u_2v_2^{-1}$ and $g$
satisfies the following recursion:
\be
\frac{g(q^2w_{12})}{g(w_{12})}=
\frac{1+qz^{-1}w_{12}}{1+zqw_{12}}
\l{rec}
\ee
We will use the following the solution of this recursion
\be
g(w_{12})=\frac{(-zqw_{12};q^2)_{\infty}}{(-z^{-1}qw_{12};q^2)_{\infty}}
\ee
Constructed $R$ - matrix differs from the Faddeev-Volkov $R$ - matrix (see
\cite{Faddeev}), by factor $f$. As it will be seen, this $R$ - matrix is
connected to the $Q$ - operator of model under consideration. To see it
let us consider some multiplication rules between the $R$ - matrix and Lax
operator. Lax operator is degenerate at the point $x_0=iq^{-1/2}$:
\be
L(x_0)=\left(\ba{c} iq^{-1/2}v^{-1} \\ u^{-1} \ea \right) \cdot \left(
\ba{cc} -ivuq^{1/2},&1 \ea \right) =\tau^-u^{-1}\rho^{-}
\ee
Introduce also
\be
\tau^+=\left(\ba{cc} 1, & -iq^{1/2}v^{-1}u \ea \right),~~\rho^+=\left(
\ba{c} 1 \\ iq^{1/2}vu \ea \right)
\ee
(it is clear that $\rho^-\rho^+=\tau^+\tau^-=0$) to construct the
projectors:
\be
\Pi^-=\tau^-(\rho^-\tau^-)^{-1}\rho^-,~~
\Pi^+=\rho^+(\tau^+\rho^+)^{-1}\tau^+
\ee
Their properties as the projectors are:
\be
(\Pi^{\pm})^2=\Pi^{\pm},~\Pi^{\pm}\Pi^{\mp}=0,~\Pi^{+}+\Pi^{-}=1
\ee
Let us also write down the expression for the operator that is inverse to
the Lax operator:
\be
\tilde{L}(x)=\left(\ba{cc} u^{-1} & -qxv^{-1} \\ qxv & u \ea
\right),~~L(x)\tilde{L}(x)=1+x^2q
\ee
It turns out that our $R$ - matrix satisfies the following triangle
conditions:
\be
\Pi^-L(x)R_{12}(-iq^{1/2}x)\Pi^+=\Pi^+R_{12}(-iq^{1/2}x)L(x)\Pi^-=0
\ee
Here the projectors act in the two dimensional auxiliary space and in the
Weyl space number 2. Moreover, the following multiplication rules hold
true:
\beq
&&\rho^-L(x)R(-iq^{1/2}x)=R(-iq^{-1/2}x)u_1 \rho^-\nonumber\\
&&L(x)R(-iq^{1/2}x)\rho^+=\rho^+R(-iq^{3/2}x)u_1^{-1}(1+qx^2) \nonumber\\
&&R(-iq^{1/2}x)L(x)\tau^-=\tau^-u_1R(-iq^{3/2}x)\nonumber\\
&&\tau^+R(-iq^{1/2}x)L(x)=u_1^{-1}(1+x^2q)R(-iq^{-1/2}x)\tau^+
\l{Rm}
\eq
For the monodromy of $R(-iq^{1/2}x)$
\be
\hat{Q}^{(1)}(x)=R_{N,a}(-iq^{1/2}x)R_{N-1,a}(-iq^{1/2}x)...R_{1,a}(-iq^{1
/2}x),
\ee
\be
tr_a\hat{Q}^{(1)}(x)=Q^{(1)}(x)
\ee
These rules guarantee the validity of Baxter equation:
\be
t(x)Q^{(1)}(x)=u_1^{-1}...u_N^{-1}(1+x^2q)^N \cdot
Q^{(1)}(q^{-1}x)+u_N...u_1 \cdot Q^{(1)}(qx)
\ee
provided the trace of $\hat{Q}^{(1)}(x)$ exists.
In our case it is sufficient to demand that the trace of permutation is
equal to the identity operator 
\be
tr_2~P_{12}=id_1.
\l{perm}
\ee
As an example let us consider the simplest case of one degree of freedom.
In this case using the property (\ref{perm}) we immediately arrive at the
following expression for $Q$ - operator:
\be
Q(i q^{-1/2}z)= 
tr_2~ R(z)=u_1^{2ln z/ln q}\sum_{k=0}(-q u_1^2)^k q^{k^2}
\frac{(z^2;q^2)_k}{(q^2;q^2)_k}
\ee
\subsection{Universal $R$ - matrix for $q$ - DST model}
Let us construct also  the universal $R$ - matrix for the $q$ - deformed
integrable DST model \cite{q-Toda}. The Hamiltonian of this model is
\be
H=\sum\limits_{n=1}^N
\left[
u_n^{-2}+a_n^+u_n^{-1} a_{n-1} u_{n-1}^{-1}\right]
\ee
where $a_n$, $a^+_n$, $u_n$ is the generators of $q$ - oscillator algebra
satisfying:
\be
au=qua,~ua^+=qa^+u,~a^+a=q^{-1}(1-u^2),~aa^+=q^{-1}(1-q^2u^2)
\ee
This Hamiltonian belongs to the family of commuting operators generated
by the transfer matrix 
\be
t(x)=Tr T(x)=Tr L_N(x)...L_1(x)
\ee
where the Lax operator for this model is given by
\be
L(x)=\left(\ba{cc} u x^{-1}+xu^{-1} & a^+ \\ a & x u \ea \right)
\ee
and is intertwined by the $R$ - matrix (\ref{trig}). In contrast to the
model, considered in the previous section, there exists the Algebraic
Bethe ansatz for $q$ - DST model (the reference state satisfies $a
\omega_0=0$). The defining equations for the universal $R$ - matrix are
(\ref{univ}). Again, it is suitable to extract the permutation operator.
The final result is
\be
R_{12}(z)=P_{12} \cdot (u_1 u_2)^{\ln z / \ln q} \cdot \sum
\limits_{k=0}^{\infty}
(a^+_2 a_1 u_1^{-1} u_2^{-1})^k (-zq^2)^k
\frac{(z^{-2};q^2)_k}{(q^2;q^2)_k}
\ee
It can be written down in the formal form similar to the expression for
universal $R$ - matrix from previous subsection
\be
R_{12}(z)=P_{12} \cdot (u_1 u_2)^{\ln z / \ln q} \frac{(-q^2 z^{-1}
w_{12};q^2)_{\infty}}{(-q^2 z w_{12};q^2)_{\infty}}
\ee
where $w_{12}=u_1^{-1}u_2^{-1}a^+_2 a_1$. Again, constructing projectors
from the Lax operator at the degeneration point (in this case $z=1$),
considering the multiplication rules (they are similar to (\ref{Rm})) and
taking the trace we arrive to the Baxter equation for the transfer-matrix
of the model and the $Q$ - operator, which is the trace of monodromy of
$R$ - matrices
\be
t(x)Q(x)=(x-x^{-1})^N Q(q^{-1}x) u_1 \dots u_N+x u_1^{-1} \dots u_N^{-1}
Q(qx)
\ee
\be
Q(x)=Tr_a R_{N,a} \dots R_{1,a}
\ee
For the $q$ - oscillator algebra  the trace is well defined
object, namely if
\be
A=\sum\limits_{k=0} \left\{ (a^+)^k A_k^{+}(u)+a^k A_k^{-}(u) \right\}
\ee
then
\be
tr~A=\sum_{n}^{\infty} A_0(q^n)
\l{trq}
\ee
but in our case it is sufficient to use the property (\ref{perm}) to
obtain the expressions for $Q$ - operators. However, in contrast to the
case of Weyl algebra, we are able to write down the permutation in the
following form 
\be
P_{12}=C_0(u_1,u_2)+\sum\limits_{n=1}^{\infty} 
\left\{(a^+_2 a_1)^k \cdot C^+_n(u_1,u_2)+(a^+_1 a_2)^k \cdot
C^-_n(u_1,u_2)\right\},
\ee
where
\beq
&&C_0=\Delta(u_1-u_2), \nonumber\\
&&C^+_n=\Delta(u_1-q^n u_2) \cdot g_2^{-1}(u_1) \dots g_2^{-1}(q^{n-1}
u_1), \nonumber\\
&&C^-_n=\Delta(u_2-q^n u_1) \cdot g_2^{-1}(u_2) \dots g_2^{-1}(q^{n-1}
u_2)
\eq
where $g_2(u)=q^{-1}(1-q^2u^2)$
and $\Delta(n_1-n_2)$ is the Kroneker symbol. It is clear that
(\ref{perm}) in the case of $q$ - oscillator algebra is in agreement with
the rule (\ref{trq}).
\section{The second $Q$ - operator}
Let us also write down the expression for the operator which is defined by
the following equations:
\be
\tilde{L}(x)=\left(\ba{cc} u^{-1} & -qxv^{-1} \\ qxv & u \ea
\right),~~L(x)\tilde{L}(x)=1+x^2q
\ee
Let us consider now the following intertwining equation:
\be
\tilde{L}(y)L(x)\tilde{R}=\tilde{R}L_1(x)\tilde{L}_2(y)
\ee
The corresponding equations are
\beq	
&&(u_1u_2^{-1}+qxyv_2^{-1}v_1) \cdot \tilde{R}=\tilde{R} \cdot (u_1
u_2^{-1}+q xyv_2v_1^{-1})\nonumber\\
&&(xu_2^{-1}v_1^{-1}-qyv_2^{-1}u_1^{-1}) \cdot \tilde{R}=\tilde{R} \cdot
(xv_1^{-1}u_2-qyu_1v_2^{-1}) \nonumber\\
&&(q y v_2 u_1-xv_1u_2) \cdot \tilde{R}=\tilde{R} \cdot (q y
v_2u_1^{-1}-xv_1v_2^{-1}) \nonumber\\
&&(u_1^{-1}u_2+qxyv_2v_1^{-1}) \cdot \tilde{R}=\tilde{R} \cdot (u_1^{-1}
u_2+q xyv_2^{-1}v_1)
\eq
It is clear that in this case $\tilde{R}=\tilde{R}(xy)$. From the first
and the last equations we see that
\beq
&&u_2^{-1}v_1^{-1} \cdot \tilde{R}=\tilde{R} \cdot v_1^{-1}u_2,~~
u_1^{-1}v_2^{-1} \cdot \tilde{R}=\tilde{R} \cdot u_1v_2^{-1} \nonumber\\
&&u_1 v_2 \cdot R=R \cdot u_1^{-1}v_2,~~ u_2 v_1 \cdot \tilde{R}=\tilde{R}
\cdot u_2^{-1}v_1
\eq
Let us look for $\tilde{R}$ in the form
\be
\tilde{R}(z)=u_2^{2lnu_1/lnq}\cdot \tilde{r}  (\tilde{w}_{12},z)
\ee
Here we denote $\tilde{w}_{12}=u_1^{-1} v_1^{-1} u_2 v_2$. Then for $f$ we
obtain:
\be
\frac{f(q^2\tilde{w}_{12},z)}{f(\tilde{w}_{12},z)}=\frac{1+qz
\tilde{w}_{12}^{-1}}
{1+qz\tilde{w}_{12}}=qz\tilde{w}_{12}^{-1} 
\cdot \frac{1+q^{-1}z^{-1} \tilde{w}_{12}}
{1+qz\tilde{w}_{12}}
\ee
The following operator function satisfies this recursion:
\be
f(\tilde{w}_{12},z)=\theta(z^{-1}q^{-2}\tilde{w}_{12}) \cdot
\frac{(-qz\tilde{w}_{12};q^2)_{\infty}}{(-q^{-1}z^{-1}\tilde{w}_{12};q^2)_
{\infty}}
\ee
Here and in the previous subsection we use standard notations for theta
function and 
$q$ -exponent:
\beq
&&\theta(\alpha)=\sum_{Z}q^{n^2}\alpha^n,~~\theta(q^2\alpha)=q^{-1}\alpha^
{-1}\theta(\alpha) \nonumber\\
&&(x;q)_n=(1-x)(1- q x)\dots(1-q^{n-1}x) \nonumber\\
&&(x;q)_{\infty}=\prod\limits_{k=0}^{\infty}(1-q^k x),~~
(1-x)(qx;q)_{\infty}=(x;q)_{\infty}
\eq
It turns out that this intertwiner has similar to (\ref{Rm}) rules of
multiplication with the Lax operator:
\beq
&&\tau^+L(x)\tilde{R}(iq^{-1/2}x)=\tilde{R}(iq^{-3/2}x)u_1 \tau^+
\nonumber\\
&&L(x)\tilde{R}(iq^{-1/2}x)\tau^-=\tau^-\tilde{R}(iq^{1/2}x)u_1^{-1}(1+qx^
2) \nonumber\\
&&\tilde{R}(iq^{-1/2}x)L(x)\rho^+=\rho^+u_1\tilde{R}(iq^{-3/2}x)
\nonumber\\
&&\rho^-\tilde{R}(iq^{-1/2}x)L(x)=u_1^{-1}(1+x^2q)\tilde{R}(iq^{1/2}x)\rho
^-
\l{Rm2}
\eq
Again, taking the trace of monodromy
$$\hat{Q}^{(2)}(x)=\tilde{R}_{N,a}(iq^{-1/2}x)\tilde{R}_{N-1,a}(iq^{-1/2}x
)...\tilde{R}_{1,a}(iq^{-1/2}x),$$ $$tr_a\hat{Q}^{(2)}(x)=Q^{(2)}(x)$$ we
arrive to the following Baxter equation:
\be
t(x)Q^{(2)}(x)=u_1^{-1}...u_N^{-1}(1+x^2q)^N \cdot
Q^{(2)}(q^{-1}x)+u_N...u_1 \cdot Q^{(2)}(qx)
\ee
However, we do not known the suitable definition of trace operation in
this case (mainly due the presence of the factor $u_2^{2lnu_1/lnq}$ in
$\tilde{R}$).

\section{Intertwining relations between $R$ and $\tilde{R}$ matricies}
The intertwiners introduced in the previous section are constructed in
particular to satisfy the following commutative relations:
\be
[t(x),Q^{(1,2)}(y)]=0
\ee
Here we get the intertwining equations between $R$ and $\tilde{R}$
matrices, that imply the mutual commutativity the corresponding $Q$ -
operators:
\be
[Q^{(1)}(x),Q^{(1)}(y)]=0,~~[Q^{(1)}(x),Q^{(2)}(y)]=0,~~[Q^{(2)}(x),Q^{(2)
}(y)]=0
\ee
Let us algebraically prove the first itntertwining relation (our proof is
similar to the proof presented in \cite{Fkash}):
\be
R_{12}(x)R_{13}(y)R_{23}(y/x)=R_{23}(y/x)R_{13}(y)R_{12}(x)
\ee
It is the Yang-Baxter equation and implies the mutual commutativity of
$Q^{(1)}$ operators. Let us write down $R$ in the following form:
\be
R_{ab}(x)=P_{ab} \cdot (u_a u_b)^{-ln x/ln q} \cdot 
\frac{(-zqw_{ab};q^2)_{\infty}}{(-z^{-1}qw_{ab};q^2)_{\infty}}=P_{ab}
\cdot 
f_{ab}(x) \cdot K(w_{ab},x)
\ee
The proof consists of four parts.
First we move permutations to the left (the products of permutations in
left and right sides are equal) and get:
\beq
&&f_{23}(x)K(w_{23},x)f_{12}(y)K(w_{12},y)f_{23}(y/x)K(w_{23},y/x)=
\nonumber\\
&&f_{12}(y/x)K(w_{12},y/x)f_{23}(y)K(w_{23},y)f_{12}(x)K(w_{12},x)
\l{int}
\eq
Second we cancel the operator-factors $f$. We notice that:
\be
w_{12}u_2^{\alpha}=q^{-\alpha}u_2^{\alpha}w_{12},~~w_{23}u_2^{\alpha}=q^{\
alpha}u_2^{\alpha}w_{23}
\ee
Using this properties let us transform (\ref{int}) to the following form:
\beq
&&f_{23}(x)f_{12}(y)f_{23}(y/x)K(w_{23}y^{-1},x)K(w_{12}y/x,y)K(w_{23},y/x
)=\nonumber\\
&&f_{12}(y/x)f_{23}(y) f_{12}(x)K(w_{12}y,y/x)K(w_{23}x^{-1},y)K(w_{12},x)
\eq
It is easy to verify that
\be
f_{23}(x)f_{12}(y)f_{23}(y/x)=f_{12}(y/x)f_{23}(y) f_{12}(x)
\ee
Finally it remains to prove the following:
\be
K(w_{23}y^{-1},x)K(w_{12}y/x,y)K(w_{23},y/x)=
K(w_{12}y,y/x)K(w_{23}x^{-1},y)K(w_{12},x)
\l{int2}
\ee
Introducing the new variables: $U=w_{23},~~V=w_{12}$ we can rewrite
(\ref{int2}) in the following form:
\be
K(U y^{-1},x)K(V y/x,y)K(U,y/x)=
K(V y,y/x)K(U x^{-1},y)K(V,x)
\l{int3}
\ee
Introduce the notation:
\be
S(U)=(-qU;q^2)_{\infty}
\ee
Making use of the threefold Jacobi product: 
$$
(y;q^2)_{\infty}(y^{-1};q^2)_{\infty}(q^2;q^2)_{\infty}=\theta_{q^2}(y)
$$
Equation (\ref{int3}) transforms to the following form
\beq
&&S((xy)^{-1}U)S(y/xU^{-1}) \theta^{-1}(x/yU) S(Vx^{-1})S(x y^{-2}V^{-1})
\cdot \nonumber\\
&&\theta^{-1}(Vx^{-1}y{2})S(Ux/y)S(x/y
U^{-1})\theta^{-1}(y/xU)=\nonumber\\
&&=S(xV)\theta^{-1}(y^2/xV)S(xy^{-2}V^{-1}) S((xy)^{-1}U)S(x
y^{-1}U^{-1})\cdot \nonumber\\
&&\theta^{-1}(xy^{-1}U)S(x^{-1}V)S( V^{-1}x^{-1})\theta^{-1}(xV)
\l{int4}
\eq
The third step is the cancellation of all theta functions.  The main
formulae in use are
\beq
&&U \theta(V)=\theta(V) q^{-1}V^{-1} U \Rightarrow f(U) \theta(V)=
\theta(V) f(q^{-1}V^{-1}U)\nonumber\\
&&U^{-1} \theta(V)=
\theta(V) VU^{-1}q^{-1} \Rightarrow f(U^{-1}) \theta(V) f(q^{-1}VU^{-1})
\eq
After proper pulling through of theta functions we get the cancellations
of all theta functions and
\beq
&&S(q^{-1}yx^{-2}UV)S(q^{-1}y^{-1}U^{-1}V^{-1})S(UV^2q^{-2}yx^{-1}) \cdot
\nonumber\\
&&S(q^2xy^{-3}V^{-2}U^{-1})S(xy^{-2}V^{-1})S(xV)=\nonumber\\
&&S(xV)S(y^{-2}xV^{-1})S((xy)^{-1}U) \cdot \nonumber\\
&&S(xy^{-1}U^{-1})S(q^{-1}yx^{-2}UV)S(q^{-1}y^{-1}U^{-1}V^{-1})
\l{int5}
\eq
On the last step we make use of the ''pentagon'' identity:
\be
S(V)S(U)=S(U)S(q^{-1}UV)S(V).
\ee
Multiple (eight times)  use of this identity leads to the fact that
(\ref{int5}) is true. The other intertwining relation may be checked
analogous way
\be
R_{23}(x/y)\tilde{R}_{12}(x)\tilde{R}_{13}(y)=\tilde{R}_{13}(y)\tilde{R}_{
12}(x)R_{23}(x/y)
\ee

\section{Conclusion}
The results obtained in the present paper are quite formal (except the
result for the universal $R$ - matrix of $q$ - DST model). To derive all
formulae we use only Weyl-type commutation relations (for generic $\alpha$
and $\beta$)
$$
u^{\alpha}v^{\beta}=q^{\alpha\beta}v^{\beta}u^{\alpha}.
$$ 
Nothing have been said about the representation of these operators. Due to
the existence of the trace operation in Weyl algebra (actually, we need
only $Tr_2~P_{12}=id_1$),
it is possible to construct at least the first $Q$ - operator connected
with the universal $R$ - matrix. Formally, the trace of monodromy of the
second intertwiner -- $\tilde{R}$ - matrix, which intertwines the Lax
operator and the inverse Lax operator, satisfies the Baxter equation, but
suitable trace procedure in this case is unknown. 
It would be interesting to see the relationship between the $M$ -
operators considered in our paper and the similar objects for the Lax
operator of Sin-Gordon model. Using  $R$ and $\tilde{R}$ matrices,
constructed in our paper  one can construct  $R$ and $\tilde{R}$ matrices
for the product of two Lax operators (\ref{Lax})
\be
L_{SG}=L_{2n,a}(kx)L_{2n-1,a}(k^{-1}x).
\l{prSG}
\ee
However to arrive to the Lax operator of SG model one should impose some
constrain on the Weyl operators
\be
u_{2n}u_{2n-1}v_{2n}v^{-1}_{2n-1}=1.
\ee
This constrain does not commutate with $R$ and $\tilde{R}$ matrices for
the product (\ref{prSG}).

This work was supported in part by grants of RFBR 00-15-96645,
01-01-00201,
01-02-16585, CRDF MO-011-0, of the Russian Ministry on the education
E00-3.3-62 and
INTAS 00-00561.

\end{document}